\documentclass[a4paper,11pt,aps,preprintnumbers,amsmath,amssymb,superscriptaddress,nofootinbib]{article}
\pdfoutput=1
\usepackage{jcappub}

\usepackage{multirow}
\usepackage{amssymb}
\usepackage{enumerate}
\usepackage{amssymb}
\usepackage{amsmath}
\usepackage{tikz}
\usepackage{feynmf}
\usepackage{makecell}
\usepackage{slashed}
\usepackage{natbib}
\usepackage{graphicx}
\hypersetup{colorlinks=true}
\usepackage{changes,color}

\usepackage{mathrsfs,amssymb}  
\usepackage{cancel}
\usepackage[normalem]{ulem}
\newcommand{\beq}{\begin{equation}}
\newcommand{\eeq}{\end{equation}}
\newcommand\bea{\begin{eqnarray}}
\newcommand\eea{\end{eqnarray}}

\newcommand{\eq}[1]{Eq.~(\ref{#1})}

\renewcommand{\l}{\left}
\renewcommand{\r}{\right}

\title{Cogenesis of baryon and lepton number asymmetries matching the EMPRESS Data} 

\author[a]{Kyu Jung Bae,}
\author[b]{Arghyajit Datta,}
\author[c]{Rinku Maji,}
\author[d]{and Wan-Il Park}
\note{\color{blue}All authors contributed equally.}
\affiliation[a]{Department of Physics, Kyungpook National University, Daegu 41566, Korea}
\affiliation[b]{
Department of Physics, Chung-Ang University, Seoul 06974, Korea}
\affiliation[c]{Cosmology, Gravity and Astroparticle Physics Group, Center for Theoretical Physics of the Universe, Institute for Basic Science, Daejeon 34126, Korea}	
\affiliation[d]{Division of Science Education and Institute of Fusion Science, Jeonbuk National University, Jeonju 54896, Korea}

\abstract{
We show that a simple supersymmetric $U(1)_{B-L}$ extension of the standard model can explain simultaneously the large electron neutrino asymmetry hinted by the recent EMPRESS data as well as the observed tiny baryon asymmetry via the resonant leptogenesis mechanism. 
 The condensation of $B-L$ Higgs dominating the universe at its decay is the sole source for these generation processes.
 Here, the infrequent decays of the $B-L$ Higgs to heavy right-handed neutrinos and successive prompt decays of these right-handed neutrinos around the electroweak phase transition produce the observed baryon asymmetry 
while the complete decay of the same $B-L$ Higgs at a later epoch leads to a large lepton number asymmetry.
The right amounts of both asymmetries are found to be obtained for the symmetry breaking scale $v_\phi \sim 10^{10}~{\rm GeV}$. 
Moreover, in a close connection to the positivity of both asymmetries, seemingly only the normal mass hierarchy of light neutrino species works.
Finally, the gravitational wave background from the topologically stable strong type-I cosmic strings, generated from the breaking of $U(1)_{B-L}$ symmetry, can be within the reach of future experiments such as ultimate DECIGO. 
 }

\emailAdd{kyujung.bae@knu.ac.kr}
\emailAdd{arghyad053@gmail.com}
\emailAdd{rinkumaji9792@gmail.com}
\emailAdd{wipark@jbnu.ac.kr}
\makeatletter
\gdef\@fpheader{}
\makeatother
\begin{document}

\maketitle

\section{Introduction}
\label{sec:intro}
It is widely recognized that our Universe is made up of matter only.
The observation of light element abundances produced during big bang nucleosynthesis (BBN)~\cite{Mossa:2020gjc,Pisanti:2020efz,Yeh:2022heq} and the anisotropies of the cosmic microwave background (CMB)~\cite{Planck:2018vyg} provide good measures of the apparent asymmetry between matter and antimatter, represented by:
\beq
Y_B^{\rm obs} \equiv \frac{\Delta n_b^{\rm obs}}{s} \simeq 8.73 \times 10^{-11},
\eeq
where $\Delta n_b \equiv n_b - n_{\bar{b}}$ with $n_b$ ($n_{\bar{b}}$) being the baryon (antibaryon) number density, and $s$ is the entropy density. The dynamical generation of such asymmetry requires new theoretical frameworks beyond the Standard Model (SM) of particle physics, as it cannot accommodate the observed value of $Y_B$.

Among many possibilities, the generation of baryon number asymmetry through leptogenesis~\cite{Fukugita:1986hr, Luty:1992un, Pilaftsis:1997jf} is arguably one of the most attractive and elegant mechanisms due to its connection with neutrino mass generation. 
Within its type-I seesaw version, the SM is conventionally extended with three heavy SM singlet right-handed neutrinos (RHN), which have a non zero Yukawa coupling with SM leptons and a Higgs doublet as well as lepton number violating Majorana mass term~\cite{Minkowski_1977,Yanagida_1979,Yanagida_1979_1,GellMann_1979,Mohapatra_1980,Schechter_1980}. 
In that case, the heaviness of the RHNs not only explains the lightness of the light SM neutrinos, but also helps generating a non zero lepton number asymmetry $Y_L$ (more specifically $B-L$ asymmetry $Y_{B-L}$) due to its CP violating out-of-equilibrium decay to (anti)leptons and Higgs. 
This lepton number asymmetry then partially converts to the baryon number asymmetry $Y_B$ via the $(B+L)$-violating sphaleron processes~\cite{tHooft:1976rip,Manton:1983nd,Klinkhamer:1984di} above the electroweak symmetry breaking scale~\cite{Arnold:1987mh,Khlebnikov:1988sr,Mottola:1988ff} leading to a relation $Y_B\sim -Y_L$. 
Without any new source of baryon number violation and sphalerons being out of equilibrium below the temperature $T_{\rm sp} \sim 137$ GeV, the $Y_B$ is then retained until the present epoch and can explain the observed baryon asymmetry of the Universe (BAU). 

Unlike the BAU, our understanding of the primordial lepton number asymmetry is limited because the asymmetry in neutrinos has never been observed.
Interestingly, the new measurement of the primordial helium abundance $Y_{\rm P} = 0.2370^{+0.0034}_{-0.0033}$ from EMPRESS 2022~\cite{Matsumoto:2022tlr} and $Y_{\rm P} = 0.2387^{+0.0036}_{-0.0031}$ from EMPRESS 2025~\cite{Yanagisawa:2025mgx} turns out to favor a very large positive asymmetry of electron neutrinos, $Y_{\nu_e}^{\rm BBN} \simeq + \mathcal{O}(10^{-4}-10^{-3})$~\cite{Burns:2022hkq, Escudero:2022okz}. 
Generating such a large value of $Y_{\nu_e}$ while not overproducing $Y_B$ requires a leptogenesis mechanism working well after the electroweak phase transition (EWPT).
Otherwise the asymmetries of neutrino flavors need to be fine-tuned such that  $\l| Y_B\r|\sim\l|Y_L\r| = \l|\sum_\alpha Y_{\nu_\alpha}\r| \ll \l| Y_{\nu_{\alpha}} \r|$ with $\alpha$ being the flavor index.
Furthermore, if one tries to explain the preferred values of ${Y_{\nu_e}}$ and $Y_B$ with the same origin, one additional difficulty arises.
A tiny $Y_B>0$ should be obtained from $Y_{B-L}>0$, which, however, gives $Y_L<0$, while EMPRESS results require $Y_{\nu_e}>0$. 
Earlier attempts have been made to generate large electron neutrino asymmetry or lepton asymmetry through the decay of non-topological solitons such as $L$-balls~\cite{Kawasaki:2002hq, Kawasaki:2022hvx} or $Q$-balls~\cite{Kasuya:2022cko}, Affleck-Dine mechanism~\cite{Casas:1997gx, Barenboim:2017dfq, Akita:2025zvq} or via resonant enhancement of CP asymmetry~\cite{Borah:2022uos, Bhandari:2023wit,ChoeJo:2024wqr,Borah:2024xoa}. 
However, most of these scenarios necessitate additional sources for baryon asymmetry or do not realize the correct sign of the neutrino asymmetry although some of scenarios such as the one in Ref.~\cite{Barenboim:2017dfq} may still work.

In this work, we propose a simple scenario of resonant leptogenesis which is responsible for $Y_{\nu_e}$ favored by the EMPRESS data. The leptogenesis works mostly below the $T_{\rm sp}$ where the sphaleron process is frozen, 
so it does not overproduce the baryon asymmetry. 
Moreover, at the price of some tuning, the same leptogenesis working partially before sphaleron decoupling can produce the correct amount of baryon asymmetry.
The key idea is (i) the possibility of a late-time production of RHNs from the decay of a long-lived scalar condensation which dominates the universe before its decay, and (ii) the effect of unequal flavor mixings of neutrinos that allows $Y_L<0$ while $Y_{\nu_e}>0$ \cite{Barenboim:2016shh} (see also Ref.~\cite{Barenboim:2016lxv} for asymmetries in the mass basis).
Note that it can be realized in a supersymmetric $U(1)_{B-L}$ extension with a very flat potential of the scalar field responsible for the mass of RHNs. 
In that case,
a large $Y_{\nu_e}$ is generated in the decay of RHNs, produced from the decay of the scalar condensation well after the EWPT, whereas the correct value of $Y_B$ can be generated from the small portion of condensation which decays to RHNs before the EWPT.
For this realization, it requires some level of cancellation between the asymmetries of neutrino flavors to achieve the correct $Y_B$.

This paper is organized as follows.
In Section~\ref{sec:cosmic-history}, we describe briefly the cosmic history associated with the scalar field responsible for the production of RHNs well after the EWPT.
In Section~\ref{sec:baryo-leptogenesis}, a model that realizes cogenesis of $Y_B^{\rm obs}$ and $Y_{\nu_e}^{\rm BBN}$ is proposed, and the details of the generation processes are described.
In Section~\ref{sec:num-analysis}, the result of the numerical analyses for the working parameter space is shown.
In Section~\ref{sec:strng-gws}, we discuss the gravitational waves expected in our scenario.
Finally, we conclude in Section~\ref{sec:conclusion}.

\section{Cosmic history: Thermal inflation and an early matter domination}
\label{sec:cosmic-history}
In the framework of supergravity which we impose implicitly as the underlying theory, a flat potential of a complex scalar field (flaton) $\phi$ around the origin, including a relevant temperature-dependent term, can have a form,

\beq
V(\phi, T) = V_0 + \l( c_T T^2 - m_{\phi,0}^2 \r) \l| \phi \r|^2 + \cdots,
\label{eq:pot}
\eeq

where $V_0$ is a constant term for the nearly vanishing cosmological constant at zero temperature, $c_T$ is a numerical constant coming from coupling of $\phi$ to particles in the thermal bath, $m_{\phi,0} \sim m_{\rm soft} \gtrsim \mathcal{O}(1)~{\rm TeV}$ with $m_{\rm soft}$ being the scale of the soft supersymmetry (SUSY) breaking mass parameters, and the ellipsis stands for term(s) to stabilize $\phi$ at low temperature.
In SUSY theories, the flaton field $\phi$ can be stabilized either by dimension six (or higher) operators or by radiative corrections to the mass-squared parameter.
In such cases, the low energy vacuum expectation value (VEV) of $\phi$, denoted as $v_\phi$, can be of an intermediate scale or higher.
When the dimension-6 operator is responsible for the stabilization, $v_\phi$ is of the form
\beq
v_\phi \gtrsim \sqrt{m_{\phi,0} M} \ ,
\eeq
where the mass parameter $M$ representing a cutoff scale may be either GUT or Planck scale, and $V_0$ is to be taken as
\beq
V_0 = c_0 m_{\phi,0}^2 v_\phi^2,
\eeq
with $c_0 = \mathcal{O}(10^{-2} - 1)$ depending on how $\phi$ is stabilized.
The physical mass of $\phi$ at the true vacuum can be expressed as
\beq
m_\phi \sim c_0^{1/2} m_{\phi,0}.
\eeq
Note that the radiative stabilization of $\phi$ is normally possible when the dimensionless couplings of $\phi$ to thermal particles are of order unity.
This implies that, if $c_0 \sim \mathcal{O}(10^{-2})$, and $c_T = \mathcal{O}(0.1-1)$.
On the other hand, if $c_0 = \mathcal{O}(1)$, which means that the stabilization of $\phi$ is achieved by higher order term(s), $c_T$ can be much smaller than unity.

\begin{figure}[t]
 \begin{center}
     \includegraphics[width=1\linewidth]{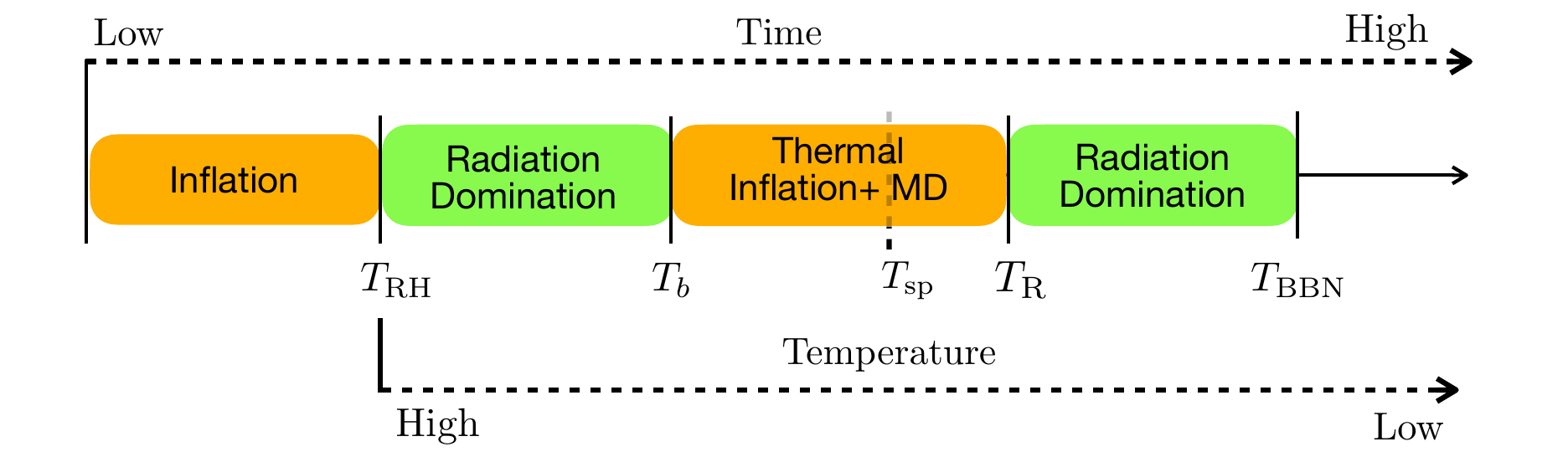}
 \end{center}
 \caption{ Schematic representation of the different phases of the Universe for our setup. Here, MD stands for matter domination.
 	}
 \label{fig:flow}
 \end{figure}
 
Fig.~\ref{fig:flow} depicts the different phases of the Universe in this case. 
With Eq.~\eqref{eq:pot} in hand,  in the very early universe for $T \gg m_{\phi,0}/c_T^{1/2}$, the Hubble-induced mass and/or thermal mass holds the flaton around the origin during and well after the primordial inflation responsible for our universe.
This leads to a thermal inflationary phase driven by the flaton VEV.
As a simple case, we may assume that the universe was dominated by radiation at the onset of this thermal inflation.\footnote{Planckian moduli whose presence is typical in SUGRA have only a minor impact on our discussion, and hence we ignore them in this work.}
Then, thermal inflation~\cite{Lazarides:1985ja,Lazarides:1986rt,Lyth:1995ka,Jeong:2004hy,Kim:2008yu} begins at $T=T_{\rm b}\sim V_0^{1/4}$ when $V_0$ starts dominating the energy density of the universe while $\phi$ is still held around the origin due to thermal effects.
It ends at $T=T_{\rm c} \approx m_{\phi,0}/c_T^{1/2}$ with $\mathcal{H}_{\rm c} \simeq V_0^{1/2}/\sqrt{3} M_{\rm P}$ as the expansion rate at the epoch.
The number of $e$-folds associated with such a single phase of thermal inflation is given by
\beq
N_e^{\rm TI} \approx \ln \l( \frac{T_{\rm b}}{T_{\rm c}} \r) \sim \ln \l( \frac{V_0^{1/4}}{m_{\phi,0}/c_T^{1/2}} \r) \sim \frac{1}{2} \ln \l( \frac{c_T c_0^{1/2}v_\phi}{m_{\rm soft}} \r),
\eeq 
where we have used $m_{\phi,0} \sim m_{\rm soft}$ in the last approximation.

For a sufficiently large $v_\phi$, it is expected to be followed by a long flaton-domination era until the flaton decays to recover the standard thermal background well before the epoch of BBN.
The decay rate of $\phi$ may be expressed as
\beq
\Gamma_\phi = \frac{\gamma_\phi}{8 \pi} \frac{m_\phi^3}{v_\phi^2},
\eeq
where $\gamma_\phi$ is a model-dependent parameter fixed in the subsequent discussion.
In this case, if $\phi$ decays dominantly to RHNs after thermal inflation, the recovery of the SM thermal bath is achieved only by subsequent decays of RHNs. For RHNs decaying right after their production from the decay of $\phi$, the maximal background temperature of the SM thermal bath after thermal inflation is found to be
\beq
T_{\rm max} \sim \l( \Gamma_\phi M_{\rm P} V_0^{1/2} \r)^{1/4} \sim \l( \frac{\gamma_\phi M_{\rm P}}{8 \pi v_\phi} \r)^{1/4} m_\phi \gg T_{\rm ew},
\eeq
where for the last inequality we have assumed $\gamma_\phi$ not very different from unity and $v_\phi \lll M_{\rm P}$ with $M_{\rm P} =2.4 \times 10^{18}~{\rm GeV}$ being the reduced Planck mass.
At a later time, temperature evolves as $T \propto a^{-3/8}$ until the epoch of reheating.
The reheating epoch in this case is defined as the epoch when the energy densities of the flaton and radiation become equal to each other.
The expansion rate at the epoch is then given by $\mathcal{H}_{\rm R} \equiv (2/5) \Gamma_\phi$, and the reheating temperature after thermal inflation is found to be 
\beq
T_{\rm R} \simeq \left( \frac{36}{5 \pi^2 g_*^{\rm R}}\right)^{1/4}\sqrt{\Gamma_\phi M_p} \ .
\label{eq:tr}
\eeq
where $g_*^{\rm R}$ is the number of relativistic degrees of freedom in the universe at the reheating epoch.
Notice that the thermal inflation dilutes the pre-existing baryon number asymmetry.
Therefore, most of baryo/leptogenesis mechanisms working before the end of thermal inflation are irrelevant, and a genesis mechanism working after thermal inflation is necessary.

If $\phi$ resides in the broken phase during the primordial inflation and the reheating after inflation is not high enough, $\phi$ could stay at the broken phase even after inflation and the associated symmetry could never be restored.
Thermal inflation cannot take place in this case, but the condensation of $\phi$ can still arise as $\phi$ start its coherent oscillations with the amplitude of $\mathcal{O}(v_\phi)$ when $\mathcal{H}_{\rm osc} \sim m_\phi$.
However, for $v_\phi$ of our interest, it is difficult for the condensation to dominate the universe at least at its decay before BBN.
In this case, the idea of generating the right amounts of baryon and lepton number asymmetries from a single source does not work, and hence we do not discuss this case any further.

\section{Cogenesis of lepton number and baryon number asymmetries}
\label{sec:baryo-leptogenesis}

We consider a thermal inflation along a D-flat direction,{\it~i.e.}, a flaton field in the SUSY theory. The subsequent decay of this flaton generates asymmetries in both baryon and lepton numbers.
This is neatly realized in a SUSY $U(1)_{B-L}$ extension of the SM such as the model considered in~\cite{Jeong:2023iei}, which is described by the superpotential,\footnote{
Depending on UV-realizations, $\mu_\Phi \Phi_1\Phi_2$ as well as $\mu H_uH_d$ (included in $W_{\rm MSSM}$) might be absent (see Ref.~\cite{Jeannerot:1998qm}).
}
\beq \label{eq:W-full}
W 
= W_{\rm MSSM} + \mu_\Phi \Phi_1 \Phi_2 + y_{N}  \Phi_1 N^2 + Y_\nu LH_u N + \Delta W_{\rm high}.
\eeq 
Here $W_{\rm MSSM}$ is  the superpotential of the minimal supersymmetric standard model (MSSM), $\Phi_1 \left( \Phi_2 \right)$ is the $U(1)_{B-L}$ Higgs with the associated charge $2 \left(-2 \right)$, $N$ is the right-handed neutrino superfield with the flavor index suppressed,  and $\Delta W_{\rm high}$ represents the leading higher order terms involving Higgs fields only:
\beq \label{eq:W-high}
\Delta W_{\rm high} =  \frac{\lambda_H}{2M} \l( H_u H_d \r)^2 + \frac{\lambda_\mu}{M} \Phi_1 \Phi_2 H_u H_d +  \frac{\lambda_\Phi}{2M} \l( \Phi_1 \Phi_2 \r)^2
\eeq 
with $M$ being a cutoff scale of the effective interactions.
Note that the $\lambda_\mu$-term can be responsible for the MSSM $\mu$-term at low energy once the $B-L$ symmetry is broken.
 In our setup, however, the MSSM $\mu$-term, $\mu H_u H_d$ already exists in $W_{\rm MSSM}$ and thus the contribution from the $B-L$ breaking can be subdominant.
For details of the model, we refer the reader to~\cite{Jeong:2023iei}.

The $U(1)_{B-L}$ $D$-flat direction consists of $\Phi_1$ and $\Phi_2$.
Denoting the direction as $\phi/\sqrt{2} = \Phi_1 = \Phi_2$, the zero-temperature potential along the direction  then can be written as
\beq \label{eq:V-flaton-susy}
V(\phi) = V_0 + m_\Phi^2 \l| \phi \r|^2 + \l( \frac{B_\Phi \mu_\Phi \phi^2}{2} + \frac{A_{\Phi}\lambda_\Phi \phi^4}{8M} + {\rm c.c.} \r) +  \l| \mu_\Phi + \frac{\lambda_\Phi \phi^2}{2 M} \r|^2 \l| \phi \r|^2,
\eeq
where $m_\Phi^2 = (m_1^2 + m_2^2)/2$ with $m_{1,2}^2$ being the soft SUSY breaking mass-square of $\Phi_{1,2}$ and $B_\Phi$ is the so-called $B$-parameter associated with the bilinear $\mu_\Phi$-term in \eq{eq:W-full}. 
Note that we have ignored any contributions involving $H_u$ and $H_d$, which appear through the $\lambda_\mu$-term in \eq{eq:W-full}.
They are negligible since $\l| \langle \lambda_\mu H_u H_d/M \rangle \r| \ll m_{\rm soft}$ at low energy with $m_{\rm soft}$ being the scale of the soft SUSY breaking mass parameters.
For simplicity, we assume $\l| B_\Phi \r| \sim \l| \mu_\Phi \r| \lesssim m_{\rm soft}$ although it depends on the mediation scenarios of the SUSY breaking.

 The flaton in this case is destabilized from the origin if there is instability in the field space of $\Phi_1$ and $\Phi_2$.
It can take place as long as $m_1^2 < 0$ at least.
Such a condition can either be achieved by a strong negative radiative running of the parameter with $y_N\sim \mathcal{O}(1)$ for at least one RHN flavor (see for example~\cite{Murayama:1992dj,Choi:1996vz,Martin:1996kn,Martin:1999hc,Bae:2014yta}) or by a sufficiently large negative value at the input scale of the theory, {\it e.g.},~\cite{Dermisek:2006qj}.
The flaton potential is stabilized by the higher order self-interaction term, {\it i.e.}, the $\lambda_{\Phi}$-term, so it has an intermediate scale VEV around $\sqrt{m_{\Phi}M/\lambda_\Phi}$.
Hence, the $\Phi_1 \Phi_2$ flat direction with its potential shown in \eq{eq:V-flaton-susy} can play the role of the flaton for thermal inflation described in the previous section.

Depending on its mass at the true vacuum, the flaton may or may not decay to electroweakinos, or the gaugino and Higginos (charginos) of $B-L$ symmetry.
If such decay channels are allowed, there is a potential danger of dark matter over-production for the stable lightest supersymmetry particle (LSP) (see, however,~\cite{Fukuda:2024ddb}).
In order to avoid such a danger, and for simplicity, we may assume that all those particles are heavier than the flaton.

The relevant part of Lagrangian for the discussion in this work can then be cast in the form\footnote{Depending on the $U(1)_{B-L}$ charge of $\Phi_1$, the mass term of RHNs could be $y_{N_i} \phi^2 \overline{N_i^c} N_i/ \Lambda$ with $\Lambda$ being a cutoff scale. Such a term is suitable for models of grand unification (see Ref.~\cite{Maji:2023fhv} for example).}
\begin{align}
	\mathcal{L}\supset y_{N_i} \phi \overline{N_i^c} N_i  
	+ (Y_\nu)_{\alpha i} \bar{\ell}_{L_\alpha} \tilde{H} N_i +h.c. ,
	\label{eq:lagrangian1}
\end{align}
where $\phi$ is the complex $U(1)_{B-L}$ Higgs which is from $\Phi_1$ in the superpotential of \eq{eq:W-full}, $N_i$ is the RHN field of the flavor $i~(=1,2,3)$, and $\tilde{H}= i \sigma_2 H^*$ with $\sigma_2$ and  $H$ being  the second Pauli matrix and the SM Higgs respectively.\footnote{
For simplicity, we used $\tilde{H}$ instead of $H_u$ even though we assume implicitly a supersymmetric UV completion. Such a simplification does not  affect the main point of our argument.}
Moreover, we assume $y_{\phi_1}\simeq y_{\phi_2} \ll y_{\phi_3}$ which makes the third RHN mass exceed those of the nearly degenerate other two and the flaton field.

Note that the thermal inflation ends as the flaton $\phi$ rolls toward its true vacuum position with a non zero VEV.
If the decay rate of the flaton, $\Gamma_\phi$, at the true vacuum is smaller than the expansion rate at the end of thermal inflation, there would be a flaton-domination era right after the thermal inflation.
In particular, if $\Gamma_\phi$ is sufficiently small so that the flaton decays mostly well after the EWPT and produces RHNs dominantly, the generation of a large electron neutrino asymmetry $Y_{\nu_e}$, favored by the EMPRESS data, may be possible without overproducing baryon asymmetry.
 It is worth noting that the flaton may decay into Higgses and Higgsinos because of $(\lambda_{\mu}/M)\Phi_1\Phi_2 H_uH_d$ coupling.
When $\phi$ takes its VEV, $\lambda_{\mu}$ induces an effective Yukawa coupling $(\lambda_{\mu}v_{\phi}/M)\phi H_uH_d$ in the superpotential, and it consequently renders the flaton decaying into Higgses and Higgsinos if kinematically allowed.
In order for the flaton to decay dominantly into RHNs, it is necessary to have $y_{N_i} = M_i/(2 v_\phi) \gg \lambda_\mu v_\phi/M$, or equivalently 
$\lambda_\mu \ll {M_i M}/({2 v_\phi^2})$, 
where $M_i$ is the mass of $N_i$.
We assume such a small $\lambda_{\mu}$ in the subsequent discussions.
The decay rate of the flaton is then given by
\beq
\Gamma_\phi= 
 \sum_{i=1,2}\frac{y_{N_i}^2}{8\pi} m_\phi \ .
\eeq
If the lighter RHNs are heavier than the SM Higgs (or up-type Higgs in the case of a SUSY model based on the MSSM), the decay rate of a RHN state $N_i$ is larger than $\mathcal{H}_{\rm EW} \sim m_{\rm EW}^2/M_{\rm P}$ with $m_{\rm EW}$ being the electroweak scale.
In this case, once produced in the decay of $\phi$ around or after the EWPT, the lighter RHNs decay promptly, forming a thermal bath consisting of the SM particles.\footnote{One may worry about the possibility of producing sneutrinos in the supersymmetric UV completion. However, sneutrinos can be heavy enough due to the soft SUSY breaking mass contribution in addition to the supersymmetric contribution from the VEV of the flaton, closing the decay channel.
We assume the case in this work.
We also assume that any of sneutrinos is not the LSP in order to avoid dark matter overproduction.}

\subsection{Lepton number asymmetry}

\begin{figure}[t]
 \begin{center}
     \includegraphics[width=0.7\linewidth]{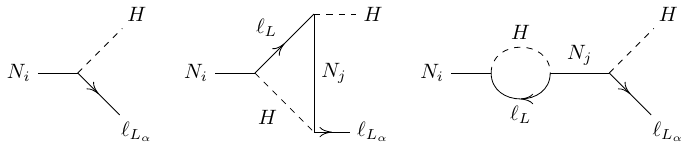}
 \end{center}
 \caption{Feynman diagrams for the RHN decays at tree as well as loop level. 
 	}
 \label{fig:feyncp}
 \end{figure}

The decay of the two lighter RHNs produces a non zero CP asymmetry along each lepton flavor $\alpha=e,~\mu,~\tau$, which is generally evaluated from the interference between the tree and one-loop diagrams comprising of vertex corrections and self energy corrections (see Fig.~\ref{fig:feyncp}). 
For a quasi-degenerate case, the self energy contribution dominates over the vertex correction~\cite{Pilaftsis:1997jf, Pilaftsis:2003gt}. Consequently,  for finite Higgs and lepton masses $m_h$ and $m_l$ respectively, the CP asymmetry takes the form of 
\begin{align}
    &\epsilon_{i \alpha}= \frac{\mathcal{F}(M_i, m_h, m_l)}{8\pi K_{ii}}
     \sum_{j\neq i}{\rm Im} \Big[(Y_{\nu})^*_{\alpha i} (Y_{\nu})_{\alpha j}\left\{K_{ij} M_i M_j +K_{ji} M_i^2 \right\}\Big]
     \frac{M_i^2-M_j^2}{(M_i^2-M_j^2)^2+M_i^2\Gamma_j^2}\label{eq:cp1}\\ &
     \text{with}\notag\\ 
     &\mathcal{F}(M_i, m_h, m_l)=
     \left(1-\frac{m_h^2-m_l^2}{M_i^2}\right)\sqrt{\left[1-\left(\frac{m_h+m_l}{M_i}\right)^2\right]
      \left[1-\left(\frac{m_h-m_l}{M_i}\right)^2\right]}.
\end{align}
Here,  $i,j=1,2~(\alpha=e,\mu,\tau)$,  $K=Y_\nu^\dagger Y_\nu$, $M_i= 2 y_{N_i} v_\phi $ is the mass of $N_i$, and $\Gamma_j= ({M_j}/({8 \pi})) K_{jj} \mathcal{F}(M_j, m_H,m_l)$ is the tree-level decay width of $N_j$. Furthermore, the phase space suppression factor $\mathcal{F}(M_i, m_H,m_l)$ arises in Eq.~\eqref{eq:cp1} for massive Higgs and leptons appearing inside the self energy loop of RHN.
Notice that the CP asymmetries in Eq.~\eqref{eq:cp1} are resonantly enhanced if $\l(M_i^2-M_j^2 \r) \simeq M_i^2 \Gamma_j^2$ which leads to an $\mathcal{O}(1)$ asymmetries when $\frac{{\rm Im}\left[K_{ij} (Y_\nu)^*_{\alpha i} (Y_\nu)_{\alpha j}\right]}{K_{ii} K_{jj}}\sim 1$ and $ \frac{{\rm Im}\left[K_{ji} (Y_\nu)^*_{\alpha i} (Y_\nu)_{\alpha j}\right]}{K_{ii} K_{jj}} \sim 1$ hold.
This resonant enhancement of the CP asymmetry along an individual lepton flavor can help us produce a large electron neutrino asymmetry.
In some cases, furthermore, we can obtain a substantially smaller total lepton asymmetry due to the cancellation among different flavors.
Therefore, it may also be possible for this setup to explain the observed baryon asymmetry via the sphaleron, as will be explored later.

The $Y_{\nu_e}$ right after the reheating depends on 
(i) the number density of the RHNs present at $T_{\rm R}$, 
(ii) the CP asymmetry generated along the electron flavor direction $\epsilon_{i(=1,2) e} $, and 
(iii) the entropy density of the Universe $s_R$ at the reheating temperature $T_{\rm R}$.  Furthermore, inverse decay of RHNs reduces the produced lepton asymmetry at $T_{\rm R}$. However, such a washout effect can be neglected by assuming $M_i \gg T_{\rm R}$ (as we will consider in our subsequent discussions). In that case, daughter Higgs and leptons will not have sufficient energy to produce back RHNs via inverse decay at $T_{\rm R}$. Note that the lepton-number-violating scattering processes may also impact the washout processes of the neutrino asymmetry. 
However, under the assumption that $M_i \gg T_{\rm R}$, the scattering rates related to these processes are also typically negligible when compared to the expansion rate of the Universe around $ T_{\rm R}$. Consequently, it is reasonable to neglect the effect of these scattering processes on lepton asymmetry generation as well. 
Finally, one can express $Y_{\nu_e}$ as
\begin{align}
	&Y_{\nu_e}^{\rm R}= \sum_{i=1}^2\epsilon_{i e} \frac{n_{N_i}^{\rm R}}{s_{\rm R}}= 6\sum_{i=1}^2\epsilon_{i e} B^\phi_i \frac{T_{\rm R}}{m_\phi}.
	\label{eq:ynu}
\end{align}
Here, we have used the relations $s_{\rm R}=\frac{4 \rho_\phi^{\rm R}}{3 T_{\rm R}}$ and $n_{N_i}^{\rm R}= \frac{8}{5}  B^\phi_i \frac{\rho_\phi^{\rm R}}{m_\phi}e^{\Gamma_\phi t_{\rm R}}$, with $B^\phi_i$ representing the branching ratio of flaton decaying to $N_{i=1,2}$ and $\Gamma_\phi t_{\rm R}=5/3$, since at reheating $\mathcal{H}_{\rm R}=\frac{2}{5}\Gamma_\phi\simeq \frac{2}{3 t_{\rm R}}$ holds where $t_{\rm R}$ is the reheating time.
Note that, with $\epsilon_{i e}\sim \mathcal{O} (1)$ and $B_i^\phi\sim 0.5$ as the best case for a large $Y_{\nu_e}$, the neutrino asymmetry depends on the ratio $T_{\rm R}/m_\phi$ that has to be greater than $ \sim 10^{-3}$ to generate the electron neutrino asymmetry favored by the EMPRESS result.

One might regard $Y_{\nu_e}$ as a constant of the evolution of the Universe, once fixed as $Y_{\nu_e}^{\rm R}$ at the reheating epoch.
However, for its subsequent evolution, the effect of neutrino flavor mixing should be taken into account.
RHN decay around $T_{\rm R}$ generally produces neutrino asymmetry along all lepton flavor directions. 
Simultaneously, because of the presence of the neutrino oscillation, neutrinos of one flavor mix with other flavors.
Consequently, the neutrino asymmetry produced along one distinct flavor can be converted to that along the other flavors before BBN. 
Furthermore, depending on the amount of initial asymmetry at $T_{\rm R}$, there can be a substantial reduction of the asymmetry for a specific neutrino flavor.  
In this work, we will mainly work with those parameters for which asymmetries satisfy 
(i)$|Y_{\nu_{\mu, \tau}}^{\rm R}| < |Y_{\nu_e}^{\rm R}|\lesssim |Y_{\nu_\mu}^{\rm R}+Y_{\nu_\tau}^{\rm R}|$, and
(ii) $Y_{\nu_e}^{\rm R}>0$, $Y_{\nu_{\mu, \tau}}^{\rm R}<0$.
It can be inferred from the earlier work~\cite{Barenboim:2016shh} that with these initial conditions the produced $Y_{\nu_e}$ at $T_{\rm R}$ would suffer a mere suppression of $Y_{\nu_e}^{\rm R}$ by a factor of about $1/3$ as the temperature of the Universe decreases to that of BBN. 
As a result, in this case the electron neutrino asymmetry survived till BBN is expected to be
\begin{align}
    Y_{\nu_e}^{\rm BBN}\approx\frac{1}{3} Y_{\nu_e}^{\rm R}.
\end{align}

\subsection{Baryon number asymmetry}
Even though the full reheating takes place well after the EWPT, RHNs are produced from decay of small fraction of $\phi$ condensation before the sphaleron process ceases at $T_{\rm sp}$.
As a result,  a small amount of lepton asymmetry can be produced from those early-produced RHNs, which is subsequently converted to the baryon asymmetry by sphalerons. 
The baryon asymmetry generated above $T_{\rm sp}$ is then diluted by the late-time entropy due to the reheating processes from $\phi$ decay.
Finally, below $T_{\rm R}$, the yield $Y_B$ is fixed, and is identified as the observed baryon asymmetry of the Universe.

The baryon asymmetry produced in this way at $T_{\rm sp}$ depends on the total lepton asymmetry (more specifically $B-L$ asymmetry) generated above $T_{\rm sp}$, which is determined by the yield of RHNs at $T_{\rm sp}$ and the CP asymmetry produced along all lepton flavor directions.  Note that in the standard radiation dominated Universe, the inverse decay and lepton number violating scattering processes can remain dominant over the Hubble expansion rate $H_{\rm sp}^{\rm RD}$ around $T_{\rm sp}$ if hierarchy between RHN mass and $T_{\rm sp}$ is not large enough, leading to a large washout of the produced lepton asymmetry. However, in our setup, Hubble expansion is dominated by the energy density of the flaton field around $T_{\rm sp}$ and can be expressed as
\begin{align}
    \mathcal{H}_{\rm sp}\equiv \mathcal{H}_{\rm sp}^{\rm MD}\approx \mathcal{H}_{\rm sp}^{\rm RD}
    \left(\frac{T_{\rm sp}}{T_{\rm R}}\right)^2, 
\end{align}
which satisfies the condition $\mathcal{H}_{\rm sp}>> \mathcal{H}_{\rm sp}^{\rm RD}$ if $T_{\rm sp}$ remains larger than $T_{\rm R}$. 
As a result, one can find a situation where the inverse decay rate as well as the scattering processes become smaller than the expansion rate (due to the domination of flaton energy density) at $T_{\rm sp}$ (as will be shown in the next section).
In such a case, solving the Boltzmann equation to evaluate the lepton number asymmetry at $T_{\rm sp}$ is not essential as analytic approximations would also yield similar result. The baryon asymmetry at $T_{\rm sp}$ then can be approximated as~\cite{Harvey:1990qw}\footnote{ If the inverse decay rate of the RHN remains subdominant compared to the Hubble expansion rate at $T_{\rm sp}$, the number density of RHN produced from the thermal bath would be suppressed compared to that from flaton decay at $T_{\rm sp}$. Consequently, the contribution to baryon asymmetry from thermal RHNs can be safely neglected.}
\begin{align}
	n_B^{\rm sp}=\frac{28}{79} n_{B-L}^{\rm sp}= -\frac{28}{79} \sum_{i,\alpha} \epsilon_{i \alpha} n_{N_i}^{\rm sp},
    \label{eq:nbsp}
\end{align}
where the number density of RHNs at $T_{\rm sp}$ can be obtained by solving the Boltzmann equation:
\begin{align}
	\dot{n}_{N_i}+3 \mathcal{H} n_{N_i}= 2\Gamma_\phi \frac{\rho_\phi}{m_\phi} B^\phi_i,
\end{align}
which approximately leads to
\begin{align}
	n_{N_i}^{\rm sp}\simeq \frac{4}{3}\frac{\rho_\phi^{\rm sp}}{m_\phi} B^\phi_i \frac{\Gamma_\phi}{\mathcal{H}_{\rm sp}} \ .
\end{align}

The expansion rate at $T_{\rm sp}$ is given by 
\beq
\mathcal{H}_{\rm sp} = \l( \frac{g_*^{\rm sp}}{25 g_*^{\rm R}} \r) \l( \frac{T_{\rm sp}}{T_{\rm R}} \r)^4 \Gamma_\phi,
\eeq
where we have used the relation $\mathcal{H}_{\rm R}=(2/5)\Gamma_{\phi}$. 

 The conversion of produced $B-L$ asymmetry into baryon asymmetry does not stop at $T_{\rm sp}$ as the sphaleron decoupling process depends on temperature of the Universe and is not instantaneous~\cite{Burnier:2005hp}. 
However, since sphaleron rate is exponentially suppressed below $T_{\rm sp}$ (as we will see in the subsequent discussion), it turns out that the baryon asymmetry production below $T_{\rm sp}$ is negligibly small.
The number density of the baryon asymmetry at $T_{\rm R}$ after thermal inflation can then be easily evaluated as
\begin{align}
	n_{B}^{\rm R}\simeq n_{B}^{\rm sp} \left(\frac{a_{\rm sp}}{a_{\rm R}}\right)^{3}=  n_{B}^{\rm sp} \left(\frac{\mathcal{H}_{\rm R}}{\mathcal{H}_{\rm sp}}\right)^{2},
\end{align}
which eventually leads to the following analytic form of the $Y_B^{\rm R}$:
\begin{align}
	Y_{B}^{\rm R}= -\frac{28}{79} \times 25 \sum_{i,\alpha} \epsilon_{i \alpha} B^\phi_i  \frac{T_{\rm R}}{m_\phi} \left(\frac{g_*^{\rm R} T_{\rm R}^4}{g_*^{\rm sp} T_{\rm sp}^4}\right) \ ,
	\label{eq:yb}
\end{align}
where $g_*^{\rm sp}$ denote the relativistic degrees of freedom present at $T_{\rm sp}$.

\section{Numerical Analyses}
\label{sec:num-analysis}

Our goal now is to understand in which parameter space we obtain the value of $Y_{\nu_e}$ favored by the EMPRESS result while keeping $Y_B$ equal to or less than the observed value.
As can be seen from Eq.~\eqref{eq:ynu}, $Y_{{\nu}_e}\sim10^{-3}$ is attainable for $T_{\rm R}/m_{\phi}\sim10^{-3}$ for $\mathcal{O}(1)$ CP asymmetry.
From Eq.~\eqref{eq:yb}, we can notice that a cancellation between the CP asymmetries of different lepton flavors, yielding a net asymmetry at the $10^{-4}$ level is necessary not to overproduce the baryon asymmetry since for $T_{\rm R}\sim 10$~GeV and $T_{\rm sp}\sim 140$ GeV, the dilution factor $\frac{g_*^{\rm R} T_{\rm R}^4}{g_*^{\rm sp} T_{\rm sp}^4}$ is of order $10^{-5}$ but $Y_B$ should be of order $10^{-10}$ or smaller.   

As discussed earlier, for quasi-degenerate RHNs the resonance condition in Eq.~\eqref{eq:cp1} can lead to a CP asymmetry of order unity, which mainly depends on the structure of the neutrino Yukawa coupling matrix $Y_{\nu}$. 
This $Y_{\nu}$ also leads to the active light neutrino mass and mixing angles via type-I seesaw mechanism. In that case, the light neutrino mass is expressed as 
\begin{align}
	m_\nu=-\frac{v^2}{2} Y_{\nu} M^{-1} Y_{\nu}^{T},
	\label{eq:type1}
\end{align}
which eventually leads to the following structure of $Y_{\nu}$ via Casas-Ibarra parametrization~\cite{Casas_2001}:
\begin{align}
Y_{\nu}=-i \frac{\sqrt{2}}{v} U D_{\sqrt{m}} {R}^{T} D_{\sqrt{M}}\,,
\label{eq:CI}
\end{align}
where $v$ is the VEV of Higgs field,\footnote{In SUSY theories, $v$ should be replaced with the VEV of the up-type Higgs denoted as $v_u$ \cite{Martin:1997ns}. In this work, we assume $v_u \approx v$.} $U$ is the Pontecorvo-Maki-Nakagawa-Sakata (PMNS) matrix which connects the flavor basis with mass basis for light neutrinos by diagonalizing the light neutrino mass matrix \eqref{eq:type1}. Here, $D_{\sqrt{m}}=  {\rm{diag}}(\sqrt{m_1},\sqrt{m_2},\sqrt{m_3})$ and $D_{\sqrt{M}}= {\rm{diag}}(\sqrt{M_1},\sqrt{M_2},\sqrt{M_3})$ represent the diagonal matrices containing the square root of light and heavy neutrino mass eigenvalues respectively, while $R$ is an orthogonal matrix 
satisfying the condition ${R}^{\rm{T}}{R}=1$.

Assuming the lightest active neutrino being massless, the $R$ matrix is parameterized by only one complex angle $\theta=\theta_R+i \theta_I$. Furthermore, depending on the mass hierarchy of the light neutrinos, the structure of the $R$ matrix changes.  For example, in the limit $M_3(\to \infty) \gg M_2\sim M_1$ and light neutrino masses following the relation $m_3 \gg m_2 > m_1(=0)$ (i.e., for normal hierarchy (NH) case) the $R$ matrix takes a form
\begin{align}
	R= 
	\begin{pmatrix}
		0& \cos\theta & \sin\theta\\
		0&-\sin\theta & \cos\theta\\
            1&0 &0
	\end{pmatrix},
\end{align}
while in the same RHN mass limit, for the inverted hierarchical (IH) scenario $m_2 > m_1 \gg m_3(=0)$, the form of the $R$ matrix becomes
\begin{align}
	R= 
	\begin{pmatrix}
		 \cos\theta & \sin\theta &0 \\
	     -\sin\theta & \cos\theta& 0\\
         0 & 0 & 1
	\end{pmatrix}.
\end{align}
 Note that for both cases, the free parameters present in the Yukawa matrix in Eq.~\eqref{eq:CI} are the complex angle $\theta$ and $M_{i=1,2}$. 
 
 For quasi-degenerate light RHNs, we now define the effective RHN mass $\overline{M}$ and mass splitting $\delta M$ as
 \begin{align}
 	\overline{M}\equiv \frac12 (M_1+M_2),~~ \delta M\equiv (M_2-M_1).
 \end{align}
For a fixed value of $\overline{M}$, the resonant enhancement of the individual CP asymmetry in Eq.~\eqref{eq:cp1} can then be observed by varying the parameter $\delta M$, $\theta_R$, and $\theta_I$. 
Moreover, additional tuning of parameters may lead to a situation where the asymmetry of the electron neutrino flavor direction becomes large enough to meet the EMPRESS result
while the asymmetries along the muon and tau directions are small and opposite in sign.
This feature can result in substantially smaller total lepton asymmetry at all temperatures which is directly connected to the correct amount of baryon asymmetry above $T_{\rm sp}$. 
 It is also worth noting that the  phase space factor $\mathcal{F}(M_i,m_h,m_l)$ in Eq.~\eqref{eq:cp1} can induce an additional suppression in CP asymmetries at high temperatures when thermal masses of Higgs and leptons dominate. 
\begin{table}[!h]
    \centering
    \begin{tabular}{|c|c|c|c|c|}
    \hline
        Temperature & $\sum_{i=1}^2 \epsilon_{i e}$ & $\sum_{i=1}^2 \epsilon_{i \mu}$ & $\sum_{i=1}^2 \epsilon_{i \tau}$  & $\sum_{i=1}^2\sum_{\alpha} \epsilon_{i \alpha}$ \\
        \hline
        \hline
         $T_{\rm R}= 8$ GeV& $+0.01508 $ & $-0.008745$ & $-0.006518$ & $-1.8 \times 10^{-4}$\\
         \hline
         $T_{\rm sp}=138$ GeV & $+0.00860$ & $-0.004990$  & $-0.003719$  &  $-1.0\times 10^{-4}$\\
         \hline
    \end{tabular}
    \caption{Numerical estimates of CP asymmetries along $e,~\mu,\tau$ directions and their sum generated at $T_{\rm R}=8$ GeV (first row)  and $T_{\rm sp}=138$ GeV (second row) for $\delta M= 7.6 \times 10^{-13}$ GeV and $\theta= -1+10^{-4} i$.}
    \label{tab:cp}
\end{table} 
Using Eq.~\eqref{eq:cp1}, the CP asymmetries along different lepton flavors and  their sum generated at $T_{\rm R}=8$ GeV (first row) and at $T_{\rm sp}=138$ GeV (second row) are enlisted in Table~\ref{tab:cp} for a fixed value of $\overline{M}=200$ GeV, $\delta M= 7.6 \times 10^{-13}$ GeV and $\theta= -1+10^{-4} i$ (within the NH setup where neutrino mass square differences and mixing angles are fixed from Ref.~\cite{Esteban:2024eli}). 
As can be seen from the last column, total CP asymmetry generated from RHN decay (for both listed temperatures) becomes suppressed due to the relative difference in sign of individual CP asymmetries. 
Furthermore, the CP asymmetries at $T_{\rm sp} = 138$ GeV undergo an additional suppression of $\mathcal{O}(0.6)$ compared to those evaluated at $T_{\rm R} = 8$ GeV due to $\mathcal{F}(M_i,m_h,m_l)$.

The specific value of $T_{\rm sp}$ is obtained by comparing the sphaleron rate for $130~\text{GeV}<T<159~\text{GeV}$~\cite{DOnofrio:2014rug} and Hubble expansion rate in matter domination as
\begin{align}
    \frac{\Gamma^{\rm sp}_{\rm broken}}{T^3}\equiv T \exp\left[0.83 \left(\frac{T}{\rm GeV}\right)-147\right]= \mathcal{H}=\left(\frac{\rho_\phi}{3 M_p^2}\right)^{1/2},
    \label{eq:sph}
\end{align}
where $\rho_\phi= \rho_\phi^{\rm R} \left({T}/{T_{R}}\right)^8$ is the energy density of the flaton field during matter domination and $M_p= 2.4 \times 10^{18}$ GeV is the reduced Planck mass. On the other hand, the specific choice of $\overline{M}=200$ GeV is made with the consideration that the decay of the RHN must be kinematically feasible at the time of $T_{\rm sp}$ and later at $T_{\rm R}$. In this case, although the zero temperature masses of SM Higgs and lepton are smaller than $\overline{M}$,  both of these particles gain thermal masses due to their interactions with the SM thermal bath. The total effective masses of Higgs and lepton can be expressed as: $m_h^2=m_{h_0}^2+ \delta m_h^2(T)$ with $m_{h_0}=125~{\rm GeV},~\delta m_h(T)\simeq 0.6~T$ GeV and $m_l(T)\simeq 0.12~ T$ GeV respectively~\cite{Weldon:1982bn,Quiros:1999jp}. Given $\overline{M}=200$ GeV, the relation $\overline{M}> m_h+m_l$ can now be maintained at $T_{\rm sp}$  as well as at $T_{\rm R}$ and one can produce non zero lepton asymmetry from RHN decay. 

\begin{figure}
 \begin{center}

     \includegraphics[width=0.45\linewidth]{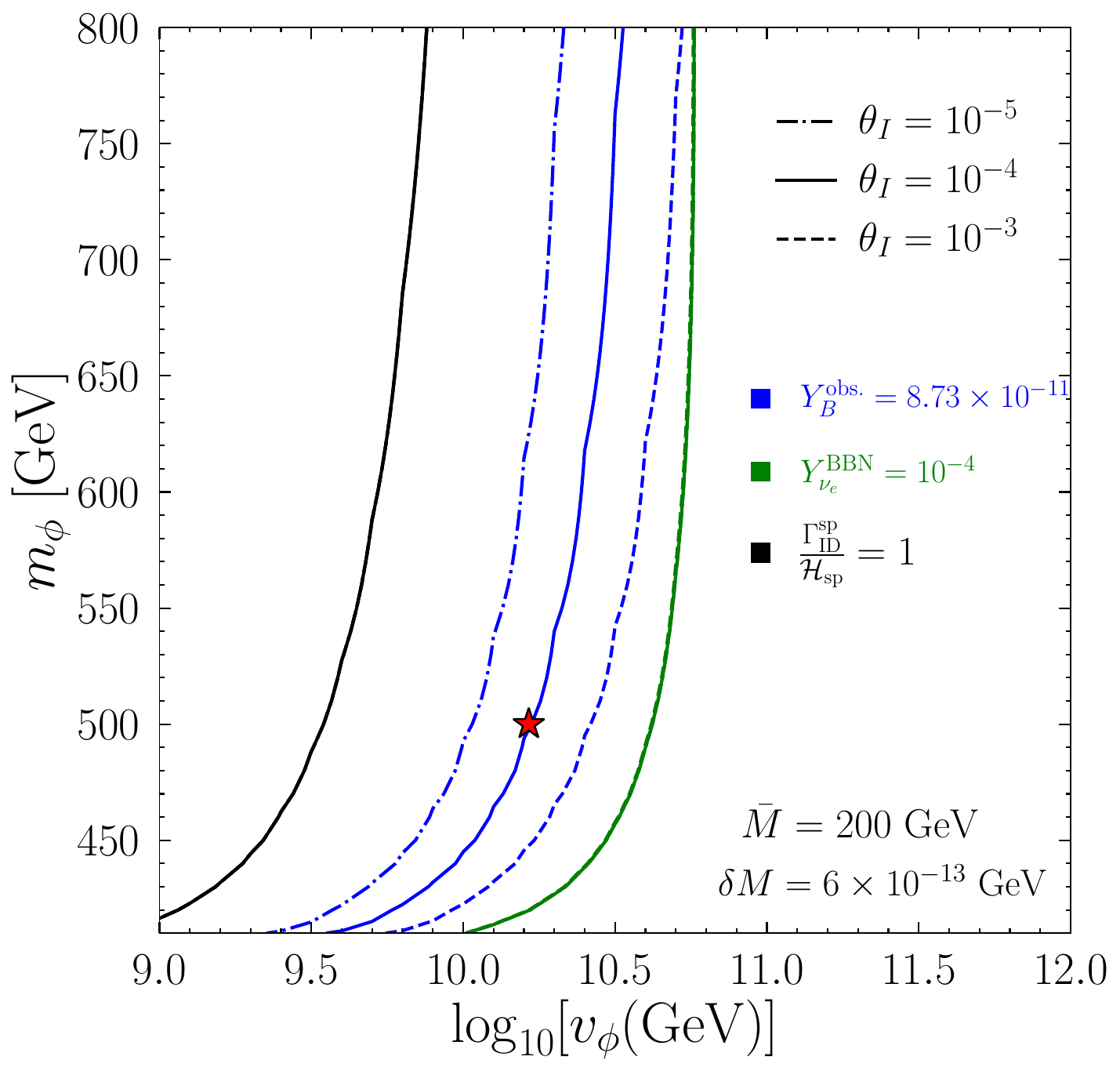}
     \includegraphics[width=0.45\linewidth]{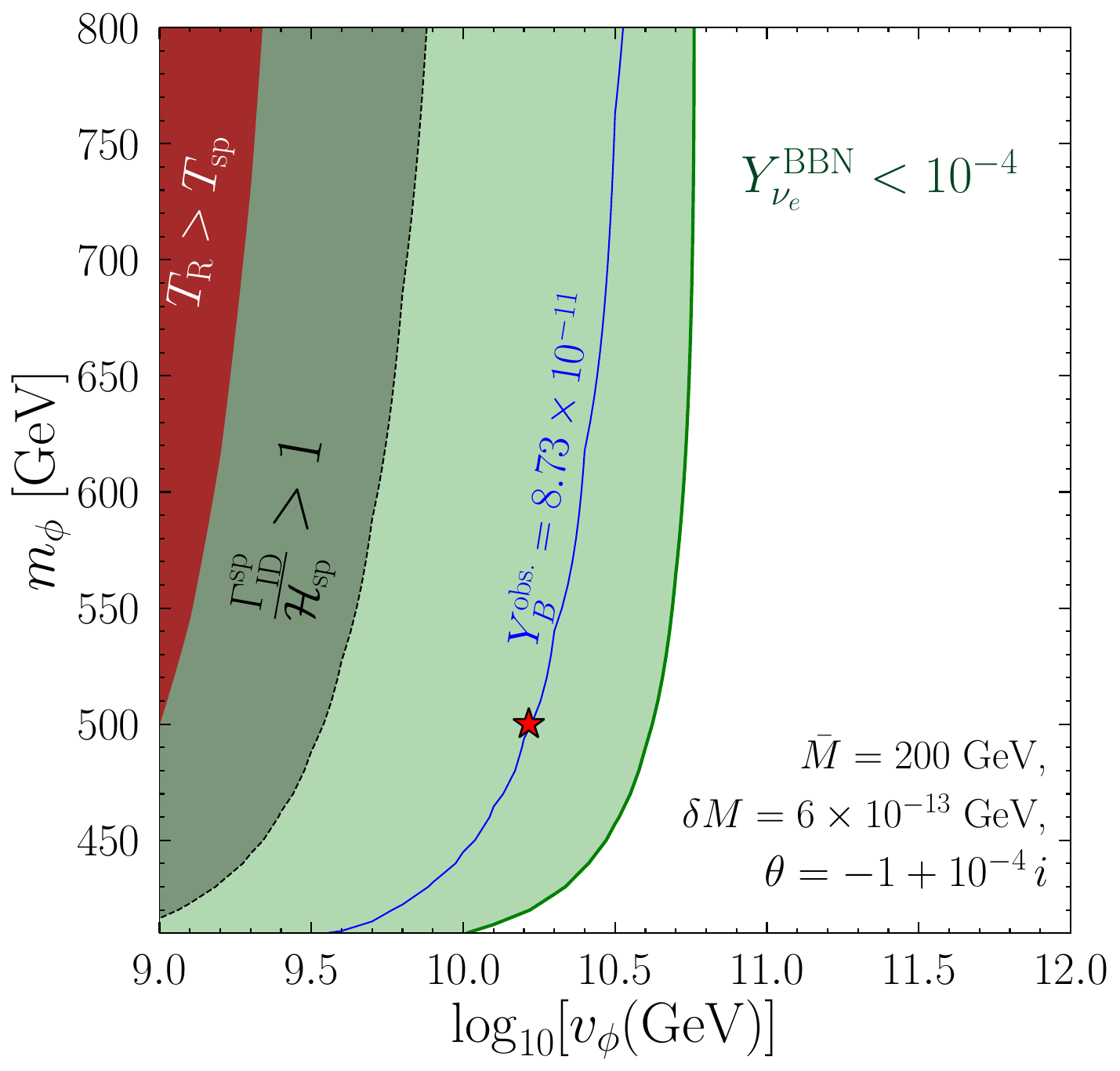}
 \end{center}
 \caption{Contour plots in $m_\phi$ vs $v_\phi$ plane indicating the parameter spaces for which the correct order of baryon asymmetry and electron neutrino asymmetry are generated for a fixed value of $\overline{M}$ and $\delta M$. In the left panel, the effect of different values of $\theta$ on $Y_{\nu_e}^{\rm BBN}=10^{-4}$, $Y_{B}^{\rm obs}=8.73 \times 10^{-11}$ and $\Gamma_{\rm ID}^{\rm sp}/\mathcal{H}_{\rm sp}=1$ contours is observed. On the other hand, right panel shows the region $Y_{\nu_e}^{\rm BBN}>10^{-4}$ (green patch), $\Gamma_{\rm ID}^{\rm sp}/\mathcal{H}_{\rm sp}>1$ (black patch) and $T_{\rm R}>T_{\rm sp}$ (red patch) for $\theta=-1+10^{-4} i$. 
}\label{fig:ybyl}
 \end{figure}

With the estimation of the CP asymmetries in hand, the produced $Y_{\nu_e}$ and  $Y_B$ at the reheating temperature can be evaluated from Eqs.~\eqref{eq:ynu} and \eqref{eq:yb} for a given set of $T_{\rm R}$, $T_{\rm sp}$ and $m_\phi$. 
Notice that the estimation of sphaleron decoupling temperature $T_{\rm sp}$ relies on the reheating temperature $T_{\rm R}$ (see Eq.~\eqref{eq:sph}), which depends on the decay rate of the flaton, evaluated in terms of $\overline{M}$ and $v_\phi$ as
\begin{align}
    \Gamma_\phi=  \sum_{i=1,2}\frac{y_{N_i}^2}{8\pi} m_\phi= \sum_{i=1,2}\left(\frac{M_i}{2v_\phi}\right)^2  \frac{m_\phi}{8\pi} \simeq \frac{\overline{M}^2}{16\pi v_\phi^2} m_\phi\,.
    \label{eq:decay}
\end{align}
 Consequently, knowledge of $v_\phi$ and $m_\phi$ is necessary to evaluate the asymmetry of the electron neutrino number and that of the baryon number at $T_{\rm R}$. 
 With $v_\phi=1.64 \times 10^{10}$ GeV and $m_\phi=500$ GeV, $T_{\rm R}$ is found to be $8$ GeV.
 In this case, the resonant leptogenesis produces a large electron neutrino asymmetry $Y_{\nu_e}^{\rm R}= 7.6\times 10^{-4}$. Using  Eq.~\eqref{eq:yb} the baryon asymmetry at the reheating temperature is found to meet the observed value, $Y_B^{\rm obs}=8.73 \times 10^{-11}$.\footnote{Following the prescription given in \cite{Burnier:2005hp}, we have numerically checked that the baryon asymmetry produced below $T_{\rm sp}$ is less than 1\% of the total baryon asymmetry. We neglect this contribution in our analyses. }
We indicate the relevant parameter point with a red star in Fig.~\ref{fig:ybyl}. 
Here, the factor ${(g_*^{\rm R} T_{\rm R}^4)}/{(g_*^{\rm sp} T_{\rm sp}^4)}=\mathcal{O}(10^{-5})$ present in Eq.~\eqref{eq:yb} plays the pivotal role (on top of the effective cancellation among CP asymmetries of $O(10^{-2})$, leading  to a total CP asymmetry of $\mathcal{O}(10^{-4})$) to reduce the baryon number asymmetry in spite of large neutrino asymmetries at $T_{\rm R}$. 
Note that for $v_\phi$, $m_\phi$ and $\overline{M}$ of our choice, the ratio between inverse decay rate for the RHNs and Hubble expansion rate $\Gamma_{\rm ID}/\mathcal{H}$ with  
$\Gamma_{\rm ID}\equiv \Gamma_{\rm ID}^{i=1,2}= \l( n_{N_i}^{\rm eq} / n_\gamma^{\rm eq} \r) \Gamma_i$, where $n_{N_i}^{\rm eq}= \frac{2T^3}{2 \pi^2} \left(\frac{M_i}{T}\right)^2 K_2\left(\frac{M_i}{T}\right)$ and $n_\gamma^{\rm eq}=(\zeta(3)/\pi^2)2 T^3$ denotes the equilibrium number density of RHNs and massless particles respectively, at $T_{\rm sp}$ and $T_{\rm R}$ can be estimated as $4 \times 10^{-2}$ and $3.24 \times 10^{-5}$ respectively. Consequently, washout effect to the produced baryon and neutrino asymmetry from inverse decay can be safely neglected.

Keeping $\overline{M}$ and $\delta M$ constant, if we now increase $\theta_I$, we observe that CP asymmetries along all the lepton flavors decrease, making the total CP asymmetry more negative. 
Consequently, the production of the observed amount of baryon number asymmetry at $T_{\rm R}$ (using Eq.~\eqref{eq:decay}) requires larger values of $v_\phi$ (when keeping $m_\phi$ constant). 
This effect can be seen in the left panel of the Fig.~\ref{fig:ybyl}, where, with gradual increase of $\theta_I$ values, $10^{-5}$ (solid curves), $10^{-4}$ (dot-dashed curves), and $10^{-3}$ (dashed curves). 
$Y_B^{\rm obs}$ contours (blue curve) shifts towards larger values of $v_\phi$. 
In the same panel, one can also notice that, since increasing $\theta_I$ from $10^{-5}$ to $10^{-3}$ does not shift the numerical estimate of electron neutrino CP asymmetry and diagonal entries of $Y_{\nu}^\dagger Y_\nu$ significantly, no substantial changes in $Y_{\nu_e}^{\rm BBN}$ (indicated by green curves) and $\Gamma_{\rm ID}^{\rm sp} / \mathcal{H_{\rm sp}}=1$ contours (black curves) are observed.

In order to see the exact dependency of $Y_B$ and $Y_{\nu_e}$ on $v_\phi$ and $m_\phi$, we now vary these two parameters keeping other parameters constant.
The allowed values of  $v_\phi$ and $m_\phi$ in this case are shown in the right panel of Fig.~\ref{fig:ybyl} where green shaded region indicates the condition $Y_{\nu_e}^{\rm BBN}> 10^{-4}$, while blue contour represents the allowed values of $v_\phi$ and $m_\phi$ for which observed baryon asymmetry can be generated.  Since larger $v_\phi$ following Eqs.\eqref{eq:tr} and \eqref{eq:decay} produces lower reheating temperature, which eventually induce smaller $Y_B$, the values of $v_\phi$ right to the blue contour underproduce the baryon asymmetry in this minimal setup. Therefore one requires an additional mechanism below $T_{\rm R}$ to generate correct amount of baryon asymmetry.
In the panel the red shaded region denotes the parameters producing $T_{\rm R} > T_{\rm sp}$ and black shaded region indicates parameters for which inverse decay of the RHN is comparable and sometimes larger than the Hubble expansion rate at the time of sphaleron decoupling, which can suppress the produced baryon asymmetry.  
Combining all these regions we find that the parameters on the blue contour satisfy all the conditions, and simultaneously explain large electron neutrino asymmetry as well as small baryon asymmetry.

Note that both baryon and neutrino asymmetries produced in this way have positive amplitudes, which were predicted by the latest EMPRESS result. 
This is one of the novel outcomes of this work. 
As already discussed, in order to generate a positive baryon asymmetry, one conventionally requires a positive $B-L$ asymmetry (or negative lepton asymmetry). 
If the same RHN decays below $T_{\rm sp}$,  one should anticipate a large negative electron neutrino asymmetry. However, as can be seen from Figure \ref{fig:ybyl}, such a conventional expectation is not always true. 
One can adjust $\overline{M}$, $\delta M$, and $\theta$ in such a way that CP asymmetry generated along the electron direction can be positive while total CP asymmetry, which is the sum of individual CP asymmetries generated along different lepton flavors, can be negative. 
Consequently, one can generate both positive baryon and electron neutrino asymmetries.

 At this stage, it is also pertinent to mention that for IH scenario, we have searched for such unique cancellation among CP asymmetry along different lepton flavor. However, due to the difference in the numerical estimates of the neutrino mass and mixing observables (compared to the case of NH), we could not find working parameters for which the same sign electron neutrino asymmetry and baryon asymmetry can be produced.

\section{Gravitational waves}
\label{sec:strng-gws}
The spontaneous breaking of the $U(1)_{B-L}$ symmetry in this setup forms topologically stable cosmic strings~\cite{Kibble:1976sj, Vilenkin:1981kz, Vilenkin:1981iu, Kibble:1982ae}.  $\phi$ is held around the origin during the thermal inflation, and the phase transition associated with the spontaneous breaking of $U(1)_{B-L}$ occurs at the end of the thermal inflation at $T=T_c$. The strings make random walk with the step of the size of order of the particle horizon at their time of formation \cite{Lazarides:130641, Vachaspati:1984dz, Vilenkin:1984ib}. These are of strong type-I, i.e., $\mathcal{R}={m_V^2}/{m_\phi^2}\gg 1$ with $m_V$ being the mass of the vector boson associated with $U(1)_{B-L}$. Furthermore, the light scalar field mediates an attractive force among the strings, which leads to the so-called zippering between the strings~\cite{Cui:2007js}. As a consequence, the string network will consist of strings with different windings.

To estimate the gravitational wave background produced from these strings, we follow the study of Ref.~\cite{Cui:2007js}.   Following Ref.~\cite{Cui:2007js} we assume that the long strings enter a scaling regime \cite{Kibble:1982cb, Kibble:1984hp, Vachaspati:1984dz, Vachaspati:1984gt, Bennett:1987vf} with their energy density being a tiny fraction ($\sim v_\phi^2/M_{\rm P}^2$) of the critical energy density of the universe at the subsequent time after their formation.
At a time $t$, the strings up to a winding number $N_w^{\rm max}$ enter the scaling regime, which can be estimated from the numerical simulation~\cite{Cui:2007js}, given as
\begin{align}
\label{eq:NwMax}
N_w^{\rm max}(t) \sim \l( \frac{t}{t_c} \r)^{0.22},
\end{align}
where $t_c\sim M_{\rm P}/\sqrt {V_0}$ is the timescale at the end of the thermal inflation. 
The average power from the network then becomes \cite{Jeong:2023iei}
\begin{align}
\label{eq:P-average}
\overline{P}_{\rm GW}(t) \xrightarrow[]{N_w^{\rm max} \gg 1} \Gamma G \mu^2(N_w^{\rm max}(t)), 
\end{align}
where $\Gamma\sim 50$ is a numerical quantity \cite{Vachaspati:1984gt, Vilenkin:2000jqa}, $G$ is the Newton's constant, and the string tension is given by \cite{Cui:2007js}
\begin{align}
\label{eq:mu-Nw}
\frac{\mu(N_w)}{2\pi v_\phi^2} 
\approx c_1 \l( 1 + c_2 \ln N_w \r),    
\end{align} 
with the numerical coefficients  
\begin{align}
\label{eq:coeff-c1}
c_1 &\approx \frac{4.2}{\ln \l(\mathcal{R} \r)} + \frac{14}{\ln^2 \l(\mathcal{R}\r)}, 
\\ \label{eq:coeff-c2}
c_2 &\approx\frac{2.6}{\ln \l(\mathcal{R} \r)} + \frac{57}{\ln^2 \l(\mathcal{R}\r)}.     
\end{align} 
 We can then express the gravitational wave background as \cite{Vachaspati:1984gt, Vilenkin:2000jqa}
\begin{align}
    \Omega_{\rm GW}(f)=\sum_{k=1}^\infty\Omega^{(k)}_{\rm GW} ,
\end{align}
where the contribution from each normal mode $k$ is given by
\begin{align}
\Omega^{(k)}_{\rm GW} 
    =& \frac{1}{\rho_c} \int_{t_F}^{t_0} d\tilde{t} \left(\frac{a(\tilde{t})}{a(t_0)}\right)^5\frac{\mathcal{F} C_{\rm eff}(t_i)}{\alpha t_i^4} \left(\frac{a(t_i)}{a(\tilde{t})}\right)^3 
     \frac{\left( 1 + c_2 \ln N_w^{\rm max}(t_i) \right)^2}{\Gamma G \mu_c + \alpha}\frac{\Gamma k^{-q}}{\zeta(q)} G\mu_c^2 \frac{2 k}{f}\theta(t_i - t_F) .
\end{align}
Here $\mu_c\equiv\mu(N_w=1)$, $\mathcal{F} \simeq 0.1$, $\alpha\simeq 0.1$, $C_{\rm eff}(t_i) = 5.7$ $(0.5)$ for the radiation (matter) dominated universe \cite{Vanchurin:2005pa, Ringeval:2005kr, Olum:2006ix, Blanco-Pillado:2013qja, Blanco-Pillado:2017oxo, Cui:2018rwi}, and $q=4/3$ for the cusp on a string loop \cite{Olmez:2010bi, Auclair:2019wcv, Cui:2019kkd, LIGOScientific:2021nrg}. 
The integration is over the time of gravitational wave radiation starting from 
\begin{align}\label{eq:tF}
    t_F=\frac{1}{\alpha}\frac{m_\phi^{-1}}{(\Gamma G\mu_c)^2},
\end{align}
which is defined as the time when the domination of the particle radiation from the string loops is over~\cite{Blanco-Pillado:1998tyu, Matsunami:2019fss, Auclair:2019jip}. The parameter $t_i$ represents the time for the formation of a loop of  initial length $l_i=\alpha t_i$ and can be derived from the frequency of the gravitational waves for a normal mode $k$ radiated at any subsequent time $\tilde{t}$ as observed today, given by
\begin{align}
    f=\frac{a(\tilde{t})}{a(t_0)}\frac{2k}{\alpha t_i -\Gamma G\mu_c(\tilde{t} - t_i)},
\end{align} 
where $a(t)$ represents the scale factor at time $t$. Notice that we have used Eq.~\eqref{eq:P-average} for $\Omega_{\rm GW}^{(k)}$. 
\begin{figure}[t]
    \centering
    \includegraphics[width=0.7\linewidth]{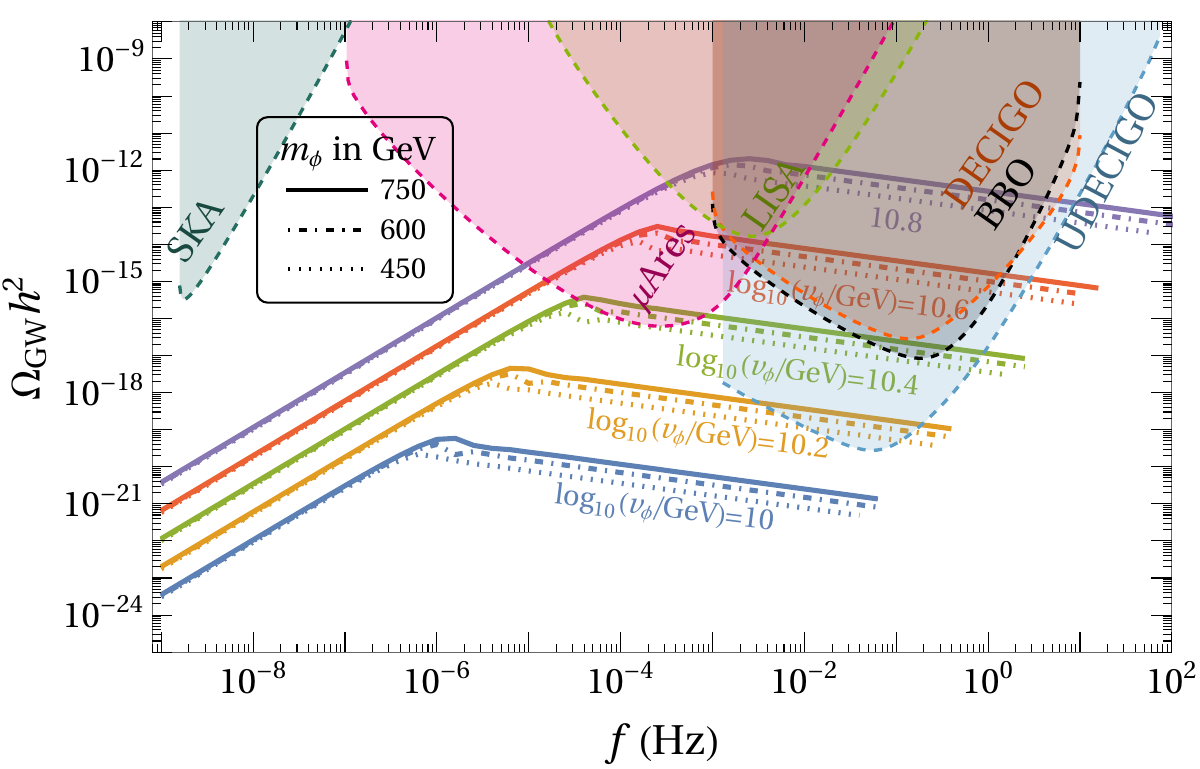}
    \caption{The gravitational wave background from the cosmic strings for $v_\phi=10^{10}$ GeV - $10^{10.8}$ GeV. For each $v_\phi$ we have shown the spectra for $m_\phi=450$ GeV (dotted), $600$ GeV (dot-dashed) and $750$ GeV (solid). The sensitivities of future detectors such as LISA \cite{Bartolo:2016ami}, DECIGO \cite{Crowder:2005nr}, ultimate DECIGO (UDECIGO) \cite{Kudoh:2005as,Kuroyanagi:2011fy}, BBO \cite{Crowder:2005nr, Corbin:2005ny}, $\mu$Ares \cite{Sesana:2019vho}, and SKA \cite{Janssen:2014dka} are depicted by their power-law integrated sensitivity curves (dashed lines). It is worth mentioning that for $v_\phi\lesssim 10^{10.5}$ GeV, we can explain both lepton and baryon asymmetries. The gravitational wave spectra for $v_\phi\gtrsim 10^{10.3}$ GeV are within the sensitivities of several proposed experiments such as LISA, $\mu$Ares, DECIGO, UDECIGO and BBO. The lepton asymmetry is satisfied for $v_\phi\gtrsim 10^{10.5}$ GeV, whereas the baryon asymmetry remains lower than the observed $Y_B^{\rm obs}=8.73\times 10^{-11}$.}
    \label{fig:GWs-strng}
\end{figure}

Figure~\ref{fig:GWs-strng} shows the expected gravitational wave background in this scenario and future observational prospects.  As can be seen from the plot, the spectra for $v_\phi\gtrsim 10^{10.3}$ GeV are within the reach of future detectors such as LISA \cite{Bartolo:2016ami}, DECIGO \cite{Komori:2025pjx}, ultimate DECIGO (UDECIGO) \cite{Kudoh:2005as,Kuroyanagi:2011fy}, BBO \cite{Crowder:2005nr, Corbin:2005ny}, and $\mu$Ares \cite{Sesana:2019vho}.
Notice also that for lower values of $v_\phi$, the gravitational waves remain suppressed due to low string tension and large string width ($\sim m_\phi^{-1}$)~\cite{Schmitz:2024hxw, Schmitz:2025uxv}.
It is also worth mentioning that the lepton asymmetry for the EMPRESS results is achievable for $10^9~\mathrm{GeV}\lesssim v_\phi\lesssim 10^{10.8}$ GeV.
The right amount of baryon and lepton asymmetries can be simultaneously obtained from the same RHN decays for the parameter space along the blue line in the right panel of Fig.~\ref{fig:ybyl}.   At this point, it is worth mentioning that the horizon-sized loops in the scaling are assumed to be relevant for the gravitational wave background which is inspired by the Nambu-Goto simulations of Refs.~\cite{Blanco-Pillado:2013qja, Blanco-Pillado:2017oxo}. In a different derivation of Ref.~\cite{Ringeval:2017eww}, a second population of smaller loops can enhanced the gravitational waves in several orders of magnitude at the higher frequencies. In addition, a large number of kinks on the loops can give rise to dominated contribution to the gravitational radiation from the kink-kink collisions in comparison with that radiated from the cusp as discussed in Ref.~\cite{LIGOScientific:2021nrg}.

Another source to the gravitational wave background can arise from a first order phase transition \cite{Yamamoto:1985rd, Easther:2008sx}. 
In that case, bubbles percolate at temperature $T_*\lesssim T_c$ at the end of thermal inflation. The dominant contribution to the gravitational waves in our scenario is from the collision of these bubbles, and in such a case, two key parameters $\alpha$ and $\beta^{-1}$ determine the characteristics of the gravitational waves~\cite{Kamionkowski:1993fg}.
The first parameter $\alpha$ denotes the ratio of the latent heat to the radiation energy density in the false vacuum, where the latent heat reflects the energy difference between the false and true vacua: 
\begin{align}
    \alpha\simeq \frac{\Delta V}{ \rho_{\rm rad}} \sim \l( \frac{v_\phi}{m_\phi} \r)^2 \ggg 1 \ .
\end{align}
The second parameter $\beta^{-1}$ is the typical size of the bubbles at the beginning of percolation. 
 
The peak amplitude of the gravitational waves from the bubble collisions is given by~\cite{Huber:2008hg}
\begin{align}
    \Omega_{\rm GW}^{\rm col}h^2\simeq 1.6\times 10^{-6}\epsilon \left(\frac{\mathcal{H}_*}{\beta}\right)^2\left(\frac{\kappa_{\rm col}\alpha}{1+\alpha}\right)^2\left(\frac{100}{g_*}\right)^{1/3},
\end{align}
at the peak frequency 
\begin{align}
    f_{\rm col}^{\rm peak} = 16.5 \times 10^{-6} \epsilon \left( \frac{0.62}{1.8 - 0.1 v_w + v_w^2} \right) \left( \frac{\beta}{\mathcal{H}_*} \right) \left( \frac{T_{\rm R}}{100\,\mathrm{GeV}} \right) \left( \frac{g_*}{100} \right)^{1/6}\,\mathrm{Hz} \ .
\end{align}
Here $\kappa_{\rm col}$ denotes the efficiency factors, $v_w \simeq 1$ is the bubble wall velocity at percolations, and $\epsilon = a(T_*)/a(T_{\rm R})\simeq (t_c/t_{\rm R})^{2/3}\sim (T_{\rm R}^2/(m_\phi v_\phi))^{2/3}\sim 10^{-7}$ accounting the dilution during the early matter domination from $t_c$ to the reheat time $t_{\rm R}\simeq \Gamma_\phi^{-1}\sim M_{\rm P}/T_{\rm R}^2$. 
Therefore, the gravitational wave background amplitude from bubble collisions is estimated to be $\Omega_{\rm GW}^{\rm col}h^2\sim 10^{-19}$ at $f_{\rm col}^{\rm peak}\sim 10^{-10}$ Hz for $\beta/\mathcal{H}_* = \mathcal{O}(10^3)$ and $\kappa_{\rm col}\sim 1$.\footnote{We use the value of $\beta/\mathcal{H}_*$ following Ref.~\cite{Easther:2008sx}.} This is out of the reach of proposed experiments such as SKA \cite{Janssen:2014dka}.

\section{Conclusions}
\label{sec:conclusion}

In this work, we have proposed a simple possibility of generating the large electron neutrino asymmetry favored by the 
recent EMPRESS results while not overproducing the baryon asymmetry.
In particular, at the price of some level of tuning, a single leptogenesis mechanism simultaneously explains the EMPRESS results
and the observed baryon asymmetry.
It is based on the resonant leptogenesis scenario, facilitating non-thermal decays of nearly degenerate heavy right-handed neutrinos from a scalar condensation which dominates the universe before its decay. 

The domination of a scalar condensation is achieved through the scenario of thermal inflation, which can be easily realized in a supersymmetric $U(1)_{B-L}$ model, for example. 
The late-time decays of the long-lived scalar condensation at a temperature well below the electroweak scale can produce the large lepton asymmetry while suppressing the baryon asymmetry.
If one tries to explain the observed baryon asymmetry from the same origin of the lepton asymmetry, the sign problem arises, which is typically expected due to $(B+L)$-violating processes.
In our scenario, it is circumvented at the price of tuning among lepton number asymmetries of all neutrino flavors.
Numerical analyses show that in order for the cogenesis to work, the symmetry breaking scale $v_\phi$ associated with the scalar condensation is constrained as $v_\phi \sim 10^{10}~{\rm GeV}$.
Also, it seems that the normal mass hierarchy of the light neutrino species is required in a close connection to the positivity of the observed baryon number asymmetry and the large electron neutrino asymmetry.

 The spontaneous breaking of $U(1)_{B-L}$ gives rise to the formation of topologically stable type-I cosmic strings. The stochastic gravitational wave background from a network of such strings can be probed in future detectors such as LISA, $\mu$Ares, DECIGO, ultimate DECIGO and BBO for $v_\phi \sim10^{10}-10^{11}~{\rm GeV}$. In addition, processes like bubble collisions during a possible first-order phase transition may produce the gravitational waves with a spectrum peaked around the frequency of a few nHz. However, the spectrum is found to be out of the reach of the proposed experiment like SKA.

\section*{Acknowledgments}

\noindent
The authors acknowledge ``IBS CTPU-CGA 2024 Summer School and Workshop for Particle Physics and Cosmology in Korea" where the discussion for the project was started. 
The work of K.J.B. was supported by the National Research Foundation of 
Korea(NRF) grant funded by the Korea government(MSIT)(No.~RS-2022-NR070836).
The work of A.D. is supported in part by Basic Science Research Program through the National Research Foundation of Korea (NRF) funded by the Ministry of Education, Science and Technology (NRF-2022R1A2C2003567).
R.M. is funded by the Institute for Basic Science under the project code: IBS-R018-D3.
The work of W.I.P. was supported by the National Research Foundation of 
Korea(NRF) grant funded by the Korea government(MSIT)(No. NRF-2022R1A4A5030362).

\bibliographystyle{JCAP.bst}
\bibliography{LL_ref.bib}

\end{document}